\documentstyle[preprint,aps]{revtex}

\begin{document}
\title{Full Fermion-Boson Vertex Function Derived in terms of the Ward-Takahashi Relations
in Abelian Gauge Theory }
\author{Han-xin He $^{a,b}$ $^{*}$ \footnotetext{$^*$ E-mail address: hxhe@iris.ciae.ac.cn}}
\address{$^{a}$ China Institute of Atomic Energy, P.O.Box 275(18), Beijing
102413, P.R.China \\
$^{b}$ Institute of Theoretical Physics, the Chinese Academy of
Science, Beijing 100080, P.R.China }

\maketitle
\begin{abstract}
I present an approach to derive the full fermion-boson vertex
function in four-dimensional Abelian gauge theory in terms of a
set of normal (longitudinal) and transverse Ward-Takahashi
relations for the fermion-boson and axial-vector vertices in
momentum space in the case of massless fermion. Such a derived
fermion-boson vertex function should be satisfied both
perturbatively and non-perturbatively. I show that, by an explicit
computation, such a derived full fermion-boson vertex function to
one-loop order leads to the same result as one obtained in
perturbation theory.

\noindent PACS number(s):  11.30.-j; 11.15.Tk.

\noindent Keywords: Full vertex function; Ward-Takahashi relations

\end{abstract}

\newpage
\section{Introduction}
It is understood that the interactions determine the structure and
properties of any theory, and the basic interactions are described
by the basic vertices. On the other hand, gauge symmetry imposes
powerful constraints on the basic vertex functions of gauge
theories, leading to the exact relations among Green's functions--
referred to as the Ward-Takahashi(WT) relations[1]. They play an
important role in providing the consistency conditions in
perturbation theory as well as in the nonperturbative study of
gauge theories through the use of the Dyson-Schwinger
equations(DSEs)[2,3]. But the normal WT identity for the vertex
specifies only its longitudinal part, leaving the transverse part
undetermined. The transverse part of the vertex has long been
known to play the crucial role in ensuring multiplicative
renormalizability of the propagator and even in determining the
propagator [2,4,5]. Therefore, in past years much effort has been
devoted to constructing the transverse part of the vertex based on
an $ansatz$ guided by perturbative constraints[2,4,5]. However,
such a constructed vertex is not unique since it is not fixed by
the symmetry of the system. The latter provides the key point in
determining the transverse part of the vertex: Like the
longitudinal part, the transverse part of the vertex should be
determined also by the WT-type constraint relations called the
transverse WT relations[6,7].

In this work, I present an approach to derive the full
fermion-boson vertex function in four-dimensional Abelian gauge
theory in terms of a set of normal and transverse WT relations for
the fermion-boson and axial-vector vertices in momentum space in
the case of massless fermion without any $ansatz$. Such a derived
fermion-boson vertex function should be satisfied both
perturbatively and non-perturbatively because it is determined
completely from the symmetry relations. This approach was proposed
but not completed yet by present author in Refs.[8,9]. This paper
presents a complete description for this approach and the complete
results for these transverse WT relations and the full
fermion-boson vertex function, as well as a proof that such a
derived full fermion-boson vertex function to one-loop order leads
to the same result as one obtained in perturbation theory.

\section{ Transverse Ward-Takahashi Relations For the Vector and
the Axial-Vector Vertex Functions}

At first, let me write the complete expressions of the normal and
transverse WT relations for the fermion-boson and axial-vector
vertices in momentum space. The normal WT identity for the
fermion-boson vertex ( i.e. vector vertex ) $\Gamma _V^\mu $ in
momentum space is well-known:
\begin{equation}
q_\mu \Gamma _V^\mu (p_1,p_2)=S_F^{-1}(p_1)-S_F^{-1}(p_2),
\end{equation}
where $q=p_1-p_2$, and $S_F$ is the full fermion propagator. This
WT identity in coordinate space is related to the divergence of
the time-ordered products of the three-point Green function
involving the vector current operator[1,10], while the transverse
WT relation for the vector vertex is related to the curl of the
time-ordered products of the three-point function involving the
vector current operator[7]:

\begin{eqnarray}
& &\partial _x^\mu \left\langle 0\left| Tj^\nu (x)\psi (x_1)\bar{\psi}(x_2)
\right| 0\right\rangle -\partial _x^\nu \left\langle 0\left|
Tj^\mu (x)\psi (x_1)\bar{\psi}(x_2)\right| 0\right\rangle \nonumber \\
&=&i\sigma ^{\mu \nu }\left\langle 0\left| T\psi (x_1)\bar{\psi}(x_2)
\right| 0\right\rangle \delta ^4(x_1-x)+i\left\langle 0\left| T\psi (x_1)
\bar{\psi}(x_2)\right| 0\right\rangle \sigma ^{\mu \nu }\delta ^4(x_2-x)\nonumber \\
& &+2m\left\langle 0\left| T\bar{\psi}(x)\sigma ^{\mu \nu }\psi (x)\psi (x_1)
\bar{\psi}(x_2)\right| 0\right\rangle \nonumber \\
& &+{\lim _{x^{\prime }\rightarrow x}}i(\partial _\lambda ^x-
\partial _\lambda ^{x^{\prime }})\varepsilon ^{\lambda \mu \nu \rho }
\left\langle 0\left| T\bar{\psi}(x^{\prime })\gamma _\rho \gamma _5
U_P (x^{\prime },x)\psi (x)\psi (x_1)\bar{\psi}(x_2)\right| 0\right\rangle ,
\end{eqnarray}
where $j^{\mu}(x)=\bar{\psi}(x)\gamma ^{\mu }\psi (x)$, and
$\sigma ^{\mu \nu }=\frac{i}{2}[\gamma ^{\mu },\gamma ^{\nu }]$.
The Wilson line $U_P (x^{\prime },x)=P\exp (-ig\int_x^{x^{\prime
}}dy^\rho A_\rho (y))$ is introduced in order that the operator be
locally gauge invariant, where $A_\mu $ are the gauge fields.
 In the QED case, $g = e$ and $A_{\rho}$ are the photon
fields.

Eq.(2) shows that the transverse part of the vector vertex is
related to the tensor and axial-vector vertices. Therefore, to
obtain complete constraint on the vector vertex, the WT relations
for the axial-vector and tensor vertices are required to build as
well. In the case of massless fermion, $m = 0$, which will be
considered in the following discussions, the tensor vertex
contribution disappears and so it is only needed to consider the
normal and transverse WT relations for the axial-vector vertex.

The normal WT identity for the axial-vector vertex function
$\Gamma _A^\mu $ in momentum space is known as[10]:
\begin{equation}
q_\mu \Gamma _A^\mu (p_1,p_2)=S_F^{-1}(p_1)\gamma _5+\gamma
_5S_F^{-1}(p_2) + i \frac{g^2}{16\pi^2} F(p_1,p_2) ,
\end{equation}
where $F(p_1,p_2) $ denotes the contribution of the axial
anomaly[11] in momentum space[9,10].

 The transverse WT relation for the
axial-vector vertex in coordinate space can be derived with the
similar procedure as that for obtaining Eq.(2), and the result
is[8,9]
\begin{eqnarray}
& &\partial _x^\mu \left\langle 0\left| Tj_5^\nu (x)\psi (x_1)\bar{\psi}(x_2)
\right| 0\right\rangle -\partial _x^\nu \left\langle 0\left| Tj_5^\mu (x)
\psi (x_1)\bar{\psi}(x_2)\right| 0\right\rangle
\nonumber \\
&=&i\sigma ^{\mu \nu }\gamma _5\left\langle 0\left| T\psi (x_1)\bar{\psi}(x_2)
\right| 0\right\rangle \delta ^4(x_1-x)-i\left\langle 0\left| T\psi (x_1)
\bar{\psi}(x_2)\right| 0\right\rangle \sigma ^{\mu \nu }\gamma _5
\delta ^4(x_2-x) \nonumber \\
& &+{\lim _{x^{\prime }\rightarrow x}}i(\partial _\lambda ^x-
\partial _\lambda ^{x^{\prime }})\varepsilon ^{\lambda \mu \nu \rho }
\left\langle 0\left| T\bar{\psi}(x^{\prime })\gamma _\rho U_P (x^{\prime },x)
\psi (x)\psi (x_1)\bar{\psi}(x_2)
\right| 0\right\rangle,
\end{eqnarray}
where $j^{\mu}_{5}(x)=\bar{\psi}(x)\gamma^{\mu}\gamma_{5}\psi(x)$.
There is no transverse axial anomaly[12].

The transverse WT relations for the vector and the axial-vector
vertex functions can be written in more clear and elegant form in
momentum space by computing the Fourier transformations of Eqs.(2)
and (4), which give
\begin{eqnarray}
& &iq^\mu \Gamma _V^\nu (p_1,p_2)-iq^\nu \Gamma _V^\mu (p_1,p_2)\nonumber \\
&=&S_F^{-1}(p_1)\sigma ^{\mu \nu }+\sigma ^{\mu \nu }S_F^{-1}(p_2)
\nonumber \\
& &+(p_{1\lambda }+p_{2\lambda })\varepsilon ^{\lambda \mu \nu \rho }
\Gamma _{A\rho }(p_1,p_2)
-\int \frac{d^{4}k}{(2 \pi)^{4}}2k_{\lambda}
\varepsilon ^{\lambda \mu \nu \rho }\Gamma _{A\rho }(p_1,p_2;k),
\end{eqnarray}
and
\begin{eqnarray}
& &iq^\mu \Gamma _A^\nu (p_1,p_2)-iq^\nu \Gamma _A^\mu (p_1,p_2)\nonumber \\
&=&S_F^{-1}(p_1)\sigma ^{\mu \nu }\gamma _5- \sigma ^{\mu \nu
}\gamma _5S_F^{-1}(p_2)\nonumber \\
& &+(p_{1\lambda }+p_{2\lambda })\varepsilon ^{\lambda \mu \nu
\rho } \Gamma _{V\rho }(p_1,p_2) -\int \frac{d^{4}k}{(2
\pi)^{4}}2k_{\lambda} \varepsilon ^{\lambda \mu \nu \rho }\Gamma
_{V\rho }(p_1,p_2;k),
\end{eqnarray}
where the integral-terms involve
 $\Gamma _{A\rho}(p_1,p_2;k)$ and
$\Gamma _{V\rho}(p_1,p_2;k)$, respectively, with the internal
momentum $k$ of the gauge boson appearing in the Wilson line.
 $\Gamma_{A\rho}(p_1,p_2;k)$ and $\Gamma _{V\rho}(p_1,p_2;k)$ are defined respectively by

\begin{eqnarray}
& &\int d^{4}x d^{4}x^{\prime}d^{4}x_{1} d^{4}x_{2}
e^{i(p_{1}\cdot x_{1} - p_{2}\cdot  x_{2} + (p_2-k)\cdot x -
(p_1-k)\cdot x^{\prime})} \langle 0|T \bar{\psi}(x^{\prime
})\gamma _\rho \gamma _5 U_P (x^{\prime },x)\psi (x)\psi
(x_1)\bar{\psi}(x_2) |0\rangle \nonumber \\
&=& (2 \pi)^{4} \delta^{4}(p_{1} - p_{2} - q) iS_{F}(p_{1})
\Gamma_{A\rho}(p_{1}, p_{2};k) iS_{F}(p_{2}) ,
\end{eqnarray}
\begin{eqnarray}
& &\int d^{4}x d^{4}x^{\prime}d^{4}x_{1} d^{4}x_{2}
e^{i(p_{1}\cdot x_{1} - p_{2}\cdot  x_{2} + (p_2-k)\cdot x -
(p_1-k)\cdot x^{\prime})} \langle 0|T \bar{\psi}(x^{\prime
})\gamma _\rho U_P (x^{\prime },x)\psi (x)\psi
(x_1)\bar{\psi}(x_2) |0\rangle \nonumber \\
&=& (2 \pi)^{4} \delta^{4}(p_{1} - p_{2} - q) iS_{F}(p_{1})
\Gamma_{V\rho}(p_{1}, p_{2};k) iS_{F}(p_{2}) ,
\end{eqnarray}
where $q=(p_1-k)-(p_2-k)$. Eqs.(7) and (8) show that $\Gamma
_{A\rho}(p_1,p_2;k)$ and $\Gamma _{V\rho}(p_1,p_2;k)$ are the
non-local axial-vector and vector vertex functions, respectively,
which and hence the integral-terms are the four-point-like
functions. These integral-terms, with relations (7) and (8), are
essential for the transverse WT relations (5) and (6) to be
satisfied both perturbatively and non-perturbatively, which were
missing in Refs.[7,8]. Indeed, as shown by Ref.[13,14], the
integral-term in Eq.[5] is crucial to prove the transverse WT
relation (5) being satisfied to one-loop order in perturbation
theory.

Eqs.(5) and (6) show that the transverse parts of the vector and
the axial-vector vertex functions are coupled each other. It
implies that the transverse parts of the vector and axial-vector
vertex functions are not independent of each other in
four-dimensional space-time.

Now there are the normal WT identities (1) and (3), which impose
the constraints on longitudinal parts of the vector and the
axial-vector vertices, respectively, and the transverse WT
relations (5) and (6), which impose the constraints on transverse
parts of these vertices.  In the case of zero fermion mass,
 Eqs.(1), (3), (5) and (6) form formally a
complete set of WT relations for the vector and the axial-vector
vertices. Then the full vector and axial-vector vertex functions
can be derived in terms of this set of WT relations.

\section{Full Fermion-Boson Vertex Function}

Now let me derive the full fermion-boson vertex ( vector vertex)
function $\Gamma _V^\mu $ by consistently solving this set of WT
relations for the vector and the axial-vector vertex functions. To
do this, multiplying both sides of Eqs.(5) and (6) by $iq_\nu $,
and then moving the terms proportional to $q_\nu \Gamma _V^\nu $
and $q_\nu \Gamma _A^\nu $ into the right-hand side of the
equations, I thus have
\begin{eqnarray}
q^2\Gamma _V^\mu (p_1,p_2) &=&q^\mu [q_\nu \Gamma _V^\nu
(p_1,p_2)]+ iS_F^{-1}(p_1)q_\nu \sigma ^{\mu \nu } + iq_\nu \sigma
^{\mu \nu }S_F^{-1}(p_2)
\nonumber \\
& &+i(p_{1\lambda }+ p_{2\lambda })q_\nu \varepsilon ^{\lambda \mu
\nu \rho }\Gamma _{A\rho }(p_1,p_2) - iq_{\nu}C_A^{\mu\nu},
\end{eqnarray}
\begin{eqnarray}
q^2\Gamma _A^\mu (p_1,p_2) &=&q^\mu [q_\nu \Gamma _A^\nu
(p_1,p_2)]+ iS_F^{-1}(p_1)q_\nu \sigma ^{\mu \nu }\gamma _5
- iq_\nu \sigma ^{\mu \nu }\gamma _5S_F^{-1}(p_2)\nonumber \\
& &+i(p_{1\lambda }+p_{2\lambda })q_\nu \varepsilon ^{\lambda \mu
\nu \rho }\Gamma _{V\rho }(p_1,p_2) - iq_{\nu}C_V^{\mu\nu},
\end{eqnarray}
where
\begin{equation}
C^{\mu\nu}_A = \int \frac{d^{4}k}{(2 \pi)^{4}}2k_{\lambda}
\varepsilon ^{\lambda \mu \nu \rho }\Gamma _{A\rho }(p_1,p_2;k),
\end{equation}
\begin{equation}
C^{\mu\nu}_V = \int \frac{d^{4}k}{(2 \pi)^{4}}2k_{\lambda}
\varepsilon ^{\lambda \mu \nu \rho }\Gamma _{V\rho }(p_1,p_2;k).
\end{equation}
Substituting Eq.(10) into Eq.(9) and using Eqs.(1) and (3), after
lengthy computations, I obtain the full fermion-boson vertex
function as follows:

\begin{equation}
\Gamma _V^\mu (p_1,p_2)=\Gamma _{V(L)}^\mu (p_1,p_2)+\Gamma _{V(T)}^\mu (p_1,p_2)
\end{equation}
with
\begin{equation}
\Gamma_{V(L)}^\mu(p_1,p_2)=q^{-2}q^\mu[S_F^{-1}(p_2)-S_F^{-1}(p_1)]
\end{equation}
and
\begin{eqnarray}
\Gamma _{V(T)}^\mu (p_1,p_2)
&=&[q^2+(p_1+p_2)^2-((p_1+p_2)\cdot q)^2q^{-2}]^{-1}\nonumber \\
& &\times \{ iS_F^{-1}(p_1)\sigma ^{\mu \nu }q_\nu
+ i\sigma ^{\mu \nu }q_\nu S_F^{-1}(p_2)\nonumber \\
& &+i[S_F^{-1}(p_1)\sigma ^{\mu \lambda }-\sigma ^{\mu \lambda }S_F^{-1}(p_2)](p_{1\lambda }+p_{2\lambda })\nonumber \\
& &+i[S_F^{-1}(p_1)\sigma ^{\lambda \nu }-\sigma ^{\lambda \nu }S_F^{-1}(p_2)]q_\nu (p_{1\lambda }+p_{2\lambda })q^\mu q^{-2}\nonumber \\
& &-i[S_F^{-1}(p_1)\sigma ^{\mu \nu }-\sigma ^{\mu \nu }S_F^{-1}(p_2)]q_\nu (p_1+p_2)\cdot qq^{-2}\nonumber \\
& & +i[S_F^{-1}(p_1)\sigma ^{\lambda \nu }+\sigma ^{\lambda \nu
}S_F^{-1}(p_2)]q_\nu (p_{1\lambda }+p_{2\lambda }) [p_1^\mu
+p_2^\mu -q^\mu (p_1+p_2)\cdot qq^{-2}]q^{-2}
\nonumber \\
& & - iq_{\nu}C^{\mu\nu}_A
+q_{\nu}q_{\alpha}q^{-2}(p_{1\lambda}+p_{2\lambda})
\varepsilon ^{\lambda \mu \nu \rho }C^{\rho\alpha}_V \nonumber \\
& & - iq_{\nu}(p_{1\lambda}+p_{2\lambda}) [p_1^\mu +p_2^\mu -q^\mu
(p_1+p_2)\cdot qq^{-2}]q^{-2}C^{\lambda\nu}_A \}.
\end{eqnarray}
Here the longitudinal part of the vertex, given by Eq.(14), is a
natural result of the normal WT relation (1), while the transverse
part of the vertex, given by Eq.(15), is derived from the
transverse WT relations for the vector and axial-vector vertices,
Eqs.(5) and (6). Consequently, the longitudinal vertex (14) as
well as the transverse vertex (15) and then the full vertex
function (13) should be satisfied both perturbatively and
nonperturbatively. In next section I will show that, by an
explicit computation, the fermion-boson vertex given by
Eqs.(13)-(15) is satisfied indeed to one-loop order in
perturbation theory.

\section{Full Fermion-Boson Vertex Function To One-loop Order}

The full fermion-boson vertex (13) consists of the longitudinal
part $\Gamma^{\mu}_{V(L)}$ and the transverse part
$\Gamma^{\mu}_{V(T)}$, where $\Gamma^{\mu}_{V(T)}$ involves two
parts of contributions from the full fermion propagator and the
four-point-like functions, which may be denoted as
$\Gamma^{\mu(I)}_{V(T)}$ and $\Gamma^{\mu (II)}_{V(T)}$ ,
respectively. To prove $\Gamma^{\mu}_{V}$ given by Eqs.(13)-(15)
being satisfied to one-loop order, it is required to calculate the
contributions from the four-point-like functions. In perturbation
theory, the four-point-like functions (7) and (8) and then the
integral-terms (11) and (12) can be calculated in the interaction
representation order by order. At one-loop order the integral-term
(11) can be written from Eq.(7) straightforwardly:

\begin{eqnarray}
& &C_A^{\mu\nu} = \int \frac{d^{4}k}{(2 \pi)^{4}}2k_{\lambda}
\varepsilon ^{\lambda \mu \nu \rho }\Gamma _{A\rho }(p_1,p_2;k)\nonumber \\
&=& g^2\int \frac{d^{4}k}{(2 \pi)^{4}}2k_{\lambda}
\varepsilon ^{\lambda \mu \nu \rho }\gamma^{\alpha}
\frac{1}{{\makebox[-0.8 mm][l]{/}{p}}_1-
{\makebox[-0.8 mm][l]{/}{k}}-m}\gamma^{\rho}\gamma_5
\frac{1}{{\makebox[-0.8 mm][l]{/}{p}}_2-{\makebox[-0.8 mm][l]{/}{k}}-m}\gamma^{\beta}
\frac{-i}{k^2}[g_{\alpha\beta}+(\xi-1)\frac{k_{\alpha}k_{\beta}}{k^2}]\nonumber \\
& &+g^2\int \frac{d^{4}k}{(2 \pi)^{4}}2
\varepsilon ^{\alpha \mu \nu \rho }[\gamma^{\beta}
\frac{1}{{\makebox[-0.8 mm][l]{/}{p}}_1-{\makebox[-0.8 mm][l]{/}{k}}-m}\gamma^{\rho}\gamma_5 + \gamma^{\rho}\gamma_5
\frac{1}{{\makebox[-0.8 mm][l]{/}{p}}_2-{\makebox[-0.8 mm][l]{/}{k}}-m}\gamma^{\beta}]
\frac{-i}{k^2}[g_{\alpha\beta}+(\xi-1)\frac{k_{\alpha}k_{\beta}}{k^2}]
,
\end{eqnarray}
where ${\makebox[-0.8 mm][l]{/}{k}}=\gamma_{\mu}k^{\mu}$, and
$\xi$ is the covariant gauge parameter. The last two terms in the
right-hand side of Eq.(16) are the one-loop self-energy
contributions accompanying the vertex correction.  Replacing
$\gamma^{\rho}\gamma_5$ by $\gamma^{\rho}$ in Eq.(16), then the
integral-term (12) at one-loop order, $C_V^{\mu\nu}$, can be
written.

The integral-term at one-loop order given by Eq.(16) was computed
in Refs.[13,14] where the transverse WT relation (5) was proved to
be satisfied to one-loop order. The result is
\begin{equation}
C^{\mu\nu}_A =- \Sigma (p_1)\sigma ^{\mu \nu } -
\sigma ^{\mu \nu }\Sigma (p_2) - Q^{\mu\nu}_V,
\end{equation}
where $\Sigma (p_i)$ ( $i=1,2$) is the one-loop fermion self-energy and

\begin{eqnarray}
Q^{\mu\nu}_V
&=& - \frac{i\alpha}{4 \pi^3}\{
\gamma_{\alpha}{\makebox[-0.8 mm][l]{/}{p}}_1
( {\makebox[-0.8 mm][l]{/}{p}}_1\sigma^{\mu \nu}+\sigma^{\mu \nu}{\makebox[-0.8 mm][l]{/}{p}}_2 ) {\makebox[-0.8 mm][l]{/}{p}}_2
\gamma^{\alpha}J^{(0)}
-\gamma_{\alpha} [ {\makebox[-0.8 mm][l]{/}{p}}_1
( {\makebox[-0.8 mm][l]{/}{p}}_1\sigma^{\mu \nu}+\sigma^{\mu \nu}{\makebox[-0.8 mm][l]{/}{p}}_2 )
\gamma^{\lambda} \nonumber \\
& &+ \gamma^{\lambda} ( {\makebox[-0.8 mm][l]{/}{p}}_1\sigma^{\mu \nu}+\sigma^{\mu \nu}{\makebox[-0.8 mm][l]{/}{p}}_2 ) {\makebox[-0.8 mm][l]{/}{p}}_2 ]\gamma^{\alpha}J^{(1)}_{\lambda}
+\gamma_{\alpha}\gamma^{\lambda}( {\makebox[-0.8 mm][l]{/}{p}}_1\sigma^{\mu \nu}+\sigma^{\mu \nu}{\makebox[-0.8 mm][l]{/}{p}}_2  )\gamma^{\eta}\gamma^{\alpha}J^{(2)}_{\lambda \eta} \nonumber \\
& &+(\xi-1)[( {\makebox[-0.8 mm][l]{/}{p}}_1\sigma^{\mu \nu}+\sigma^{\mu \nu}{\makebox[-0.8 mm][l]{/}{p}}_2 )K^{(0)}
- ( p_1^2 \gamma^{\lambda}\sigma^{\mu \nu}
+ p_2^2\sigma^{\mu \nu}\gamma^{\lambda}
+ {\makebox[-0.8 mm][l]{/}{p}}_1\sigma^{\mu \nu}{\makebox[-0.8 mm][l]{/}{p}}_2 \gamma^{\lambda} \nonumber \\
& &+ \gamma^{\lambda}{\makebox[-0.8 mm][l]{/}{p}}_1\sigma^{\mu
\nu}{\makebox[-0.8 mm][l]{/}{p}}_2 ) J^{(1)}_{\lambda} +
\gamma^{\lambda}( p_1^2 \sigma^{\mu\nu}{\makebox[-0.8
mm][l]{/}{p}}_2 + p_2^2 {\makebox[-0.8 mm][l]{/}{p}}_1 \sigma^{\mu
\nu} )\gamma ^{\eta} I^{(2)}_{\lambda \eta}]\}.
\end{eqnarray}
Here $\alpha=g^2/4\pi$, $J^{(0)}$, $J^{(1)}_{\lambda}$,
 $J^{(2)}_{\lambda \eta}$, $K^{(0)}$ and $I^{(2)}_{\lambda \eta}$ are some
 integrals:
 \begin{equation}
 J^{(0)} = \int_M d^{4}k\frac{1}{k^2[( p_1-k )^2 + i\epsilon]
[( p_2-k )^2 + i\epsilon]} ,
\end{equation}
\begin{equation}
 K^{(0)} = \int_M d^{4}k\frac{1}{[( p_1-k )^2 + i\epsilon]
 [( p_2-k )^2 + i\epsilon] } ,
\end{equation}
\begin{equation}
 I^{(2)}_{\lambda \eta} = \int_M d^{4}k\frac{k_{\lambda}k_{\eta}}
{k^4 [( p_1-k )^2 + i\epsilon][(p_2-k )^2 + i\epsilon]} .
\end{equation}
Replacing 1 by $k_{\lambda}$ and  $k_{\lambda}k_{\eta}$ in the
numerator of Eq.(19) gives the expressions of $J^{(1)}_{\lambda}$
and $J^{(2)}_{\lambda \eta}$, respectively. These integrals can be
carried out in the cutoff regularization scheme or in the
dimensional regularization scheme[4,5] .

The integral-term (12) at one-loop order can be computed
similarly, which gives

\begin{equation}
C^{\mu\nu}_V =- \Sigma (p_1)\sigma ^{\mu \nu }\gamma_5 +
\sigma ^{\mu \nu }\gamma_5\Sigma (p_2) - Q^{\mu\nu}_A,
\end{equation}
where
\begin{eqnarray}
 Q^{\mu\nu}_A
&=& - \frac{i\alpha}{4 \pi^3}\{ \gamma_{\alpha}{\makebox[-0.8
mm][l]{/}{p}}_1 ( {\makebox[-0.8 mm][l]{/}{p}}_1\sigma^{\mu
\nu}+\sigma^{\mu \nu}{\makebox[-0.8 mm][l]{/}{p}}_2 )\gamma_5
{\makebox[-0.8 mm][l]{/}{p}}_2 \gamma^{\alpha}J^{(0)}
-\gamma_{\alpha} [ {\makebox[-0.8 mm][l]{/}{p}}_1 ( {\makebox[-0.8
mm][l]{/}{p}}_1\sigma^{\mu \nu}+\sigma^{\mu \nu}{\makebox[-0.8
mm][l]{/}{p}}_2 )\gamma_5
\gamma^{\lambda} \nonumber \\
& &+ \gamma^{\lambda} ( {\makebox[-0.8 mm][l]{/}{p}}_1\sigma^{\mu
\nu}+\sigma^{\mu \nu}{\makebox[-0.8 mm][l]{/}{p}}_2 )
\gamma_5{\makebox[-0.8 mm][l]{/}{p}}_2
]\gamma^{\alpha}J^{(1)}_{\lambda}
+\gamma_{\alpha}\gamma^{\lambda}( {\makebox[-0.8 mm][l]{/}{p}}_1\sigma^{\mu \nu}+\sigma^{\mu \nu}{\makebox[-0.8 mm][l]{/}{p}}_2  )\gamma_5\gamma^{f}\gamma^{\alpha}J^{(2)}_{\lambda f} \nonumber \\
& &+(\xi-1)[( {\makebox[-0.8 mm][l]{/}{p}}_1\sigma^{\mu
\nu}+\sigma^{\mu \nu}{\makebox[-0.8 mm][l]{/}{p}}_2
)\gamma_5K^{(0)} - ( p_1^2 \gamma^{\lambda}\sigma^{\mu
\nu}\gamma_5 - p_2^2\sigma^{\mu \nu}\gamma_5\gamma^{\lambda}
+ {\makebox[-0.8 mm][l]{/}{p}}_1\sigma^{\mu \nu}\gamma_5{\makebox[-0.8 mm][l]{/}{p}}_2 \gamma^{\lambda} \nonumber \\
& &- \gamma^{\lambda}{\makebox[-0.8 mm][l]{/}{p}}_1\sigma^{\mu
\nu}\gamma_5{\makebox[-0.8 mm][l]{/}{p}}_2 ) J^{(1)}_{\lambda} +
\gamma^{\lambda}( p_1^2 \sigma^{\mu\nu}\gamma_5{\makebox[-0.8
mm][l]{/}{p}}_2 - p_2^2 {\makebox[-0.8 mm][l]{/}{p}}_1 \sigma^{\mu
\nu}\gamma_5 )\gamma ^{f} I^{(2)}_{\lambda f}]\}.
\end{eqnarray}

Now substituting the fermion propagator to one-loop order
$S^{-1}_F(p_i) = {\makebox[-0.8 mm][l]{/}{p}}_i - \Sigma(p_i), \ \
\ i=1,2,$ together with Eqs.(17)-(18) and Eqs.(22)-(23) into
Eqs.(13)-(15), after some algebraic calculations, I obtain
\begin{equation}
 \Gamma^{\mu}_V = \Gamma^{\mu}_{V(L)} + \Gamma^{\mu}_{V(T)}
 = \gamma^{\mu} + \Lambda^{\mu}_V ,
\end{equation}
where
\begin{eqnarray}
 \Lambda_V^{\mu}(p_1,p_2)
&=& - \frac{i\alpha}{4 \pi^3}\{
\gamma^{\alpha}{\makebox[-0.8 mm][l]{/}{p}}_1\gamma^{\mu}{\makebox[-0.8 mm][l]{/}{p}}_2
\gamma_{\alpha}J^{(0)} \nonumber \\
& &- (\gamma^{\alpha}{\makebox[-0.8
mm][l]{/}{p}}_1\gamma^{\mu}\gamma^{\lambda}\gamma_{\alpha}
+\gamma^{\alpha}\gamma^{\lambda}\gamma^{\mu}{\makebox[-0.8
mm][l]{/}{p}}_2\gamma_{\alpha}) J^{(1)}_{\lambda}
+\gamma^{\alpha}\gamma^{\lambda}\gamma^{\mu}\gamma^{f}\gamma_{\alpha}
J^{(2)}_{\lambda f} \nonumber \\
& &+ (\xi - 1)[ \gamma^{\mu} K^{(0)}
- (\gamma^{\lambda}{\makebox[-0.8 mm][l]{/}{p}}_1 \gamma^{\mu} +
 \gamma^{\mu}{\makebox[-0.8 mm][l]{/}{p}}_2\gamma^{\lambda} )J^{(1)}_{\lambda}
+\gamma^{\lambda}{\makebox[-0.8 mm][l]{/}{p}}_1\gamma^{\mu}{\makebox[-0.8 mm][l]{/}{p}}_2\gamma^f I^{(2)}_{\lambda f}]
\} ,
\end{eqnarray}
which is the familiar expression of one-loop vector vertex
function in perturbation theory[4,5]. This shows that the full
fermion-boson vertex function given by Eqs.(13)-(15) to one-loop
order leads to the same result as one obtained in perturbation
theory. Since the four-point-like functions,
$\Gamma_{A\rho}(p_1,p_2;k)$ and $\Gamma_{V\rho}(p_1,p_2;k)$, can
be calculated order by order in perturbation theory, thus one may
demonstrate that the full fermion-boson vertex function derived
from symmetry relations should be satisfied order by order in
perturbation theory.

\section{Conclusion and Remark}

 In this paper, I have derived the full fermion-boson vertex function
in four-dimensional Abelian gauge theory in terms of a set of
normal and transverse WT relations for the vector and axial-vector
vertices in momentum space in the case of massless fermion. The
longitudinal part of the vertex is specified by the normal WT
relation in terms of the fermion propagator, while the transverse
part of the vertex is determined by the transverse WT relations
for the vector and the axial-vector vertices. Such a derived full
fermion-boson vertex function, in principle, should be satisfied
both perturbatively and nonperturbatively because it is determined
by symmetry relations which are satisfied perturbatively and
nonperturbatively. I have shown that, by an explicit computation,
such a full fermion-boson vertex is satisfied indeed to one-loop
order in perturbation theory. By the parallel procedure, the full
axial-vector vertex function can be derived also. Thus this
provides an approach to derive the basic vertex functions from the
symmetry relations.

Notice that the transverse part of the vertex,
$\Gamma^{\mu}_{V(T)}$, derived this way separates naturally two
parts:  $\Gamma^{\mu(I)}_{V(T)}$ expressed in terms of the full
fermion propagator, and  $\Gamma^{\mu (II)}_{V(T)}$ related to the
four-point-like functions. Neglecting $\Gamma^{\mu (II)}_{V(T)}$
corresponds to the cutoff of the four-point-like functions. In
this case, it is easy to check that $\Gamma _{V(L)}^\mu (p_1,p_2)
+\Gamma _{V(T)}^{\mu(I)} (p_1,p_2) \rightarrow \gamma^{\mu}$ when
the fermion propagator is taken as bare one. It shows that $\Gamma
_{V(T)}^{\mu(I)}$ gives the leading contribution to the transverse
part of the vertex. The contribution related to the
four-point-like functions, $\Gamma^{\mu (II)}_{V(T)}$, needs to be
studied further.

$Additionally$, I checked the transverse WT relation for the
fermion-boson vertex to one-loop order in d-dimensions, with $d=4
+ \epsilon$. The result shows that this relation in d-dimensions
has the same form as one, Eq.(5) with Eq.(16), given in
4-dimensions and so there is no need for an additional piece $\sim
(d-4)$ to include for this relation to hold in 4-dimensions. Thus
it can be checked that the full fermion-boson vertex function
expressed by Eqs.(13) to (15) holds exactly in 4-dimensions.
Recently, Ref.[15] made a good comment on the potential of the
transverse WT relation to determine the full fermion-boson vertex
and then checked the transverse WT relation to one loop order in
d-dimensions. However, the authors of Ref.[15] separated out a
so-called additional piece $\sim (d-4)$ from the integral-term by
defining a modifying integral-term, which in fact does not change
the original formula of the transverse WT relation to one-loop
order. Thus it needs not have introduced such an additional piece.
The detailed discussions will be given in a separate paper[16].

\section*{Acknowledgments}

I wish to thank Y.Takahashi and F.C.Khanna for useful
conversations, thank H.S.Zong for telling me Ref.[15] and useful
conversations. This work is supported by the National Natural
Science Foundation of China under grant No.90303006.

\end{document}